\documentclass{aa}

\usepackage{natbib}                     
\usepackage{graphicx}                   
\usepackage[varg]{txfonts}              
\usepackage{hyperref}             
\hypersetup{colorlinks=true,urlcolor=blue, linkcolor=blue,  citecolor=blue}

\begin{document}

  \title{Density discrepancy between transit-timing variations and radial velocity: Insights from the host star composition}

  \author{V.~Adibekyan\inst{1,2} \and
          S.~G.~Sousa \inst{1} \and 
          E.~Delgado~Mena \inst{1} \and
          N.~C.~Santos \inst{1,2} \and \\
          G.~Israelian \inst{3,4} \and
          S.~C.~C.~Barros \inst{1,2} \and
          Zh.~Martirosyan \inst{5} \and
          A.~A.~Hakobyan\inst{6} 
 }

  \institute{
          Instituto de Astrof\'isica e Ci\^encias do Espa\c{c}o, Universidade do Porto, CAUP, Rua das Estrelas, 4150-762 Porto, Portugal \\
          \email{vadibekyan@astro.up.pt} \and
          Departamento de F\'{\i}sica e Astronomia, Faculdade de Ci\^encias, Universidade do Porto, Rua do Campo  Alegre, 4169-007 Porto, Portugal \and
          Instituto de Astrof\'{i}sica de Canarias, E-38205 La Laguna, Tenerife, Spain \and           Departamento de Astrof\`{i}sica, Universidad de La Laguna, E-38206 La Laguna, Tenerife, Spain \and
          Department of Astrophysics and General Physics, Yerevan State University, A. Manukyan str. 1, 0025 Yerevan, Armenia \and
          Center for Cosmology and Astrophysics, Alikhanian National Science Laboratory, 2 Alikhanian Brothers Str., 0036 Yerevan, Armenia
          }

  \date{Received date / Accepted date }
  \abstract
  {The determination of planetary densities from the masses derived with the radial velocity (RV) and transit-timing variation (TTV) methods reveals discrepancies. Specifically, planets detected through RV exhibit higher densities than those detected through TTV, even though their radii are similar. Understanding the origins of these discrepancies is crucial and timely, especially with upcoming ground- and space-based missions dedicated to exoplanet research.}
  {We explored the possibility that the discrepant mass/densities in the TTV and RV populations might be linked to the properties of the environments in which these planets are formed.}
  {For the largest currently available sample of FGK-type stars hosting low-mass TTV and RV planets, we determined the host star abundances. Then, by employing a simple stoichiometric model, we used these abundances to estimate the iron-to-silicate mass fraction ($f_{\mathrm{iron}}$) and the water-mass fraction ($w_{f}$) of the protoplanetary disks. We also calculated the kinematic properties of the host stars.}
  {We observed an indication that the hosts of TTV planets have slightly higher $f_{\mathrm{iron}}$ and lower $w_{f}$ values than their RV counterparts. This suggests that TTV planets (without considering their atmospheres) are denser than RV planets on average, which implies that larger atmospheres on TTV planets are required to account for their overall lower densities. However, we also note differences in the properties of the planets, such as their orbital periods, and variations in the quality of the spectroscopic data, which may have an impact on these results.}
  {Exploring the TTV-RV mass and/or density discrepancy through a chemical analysis of the host star holds promise for future research, particularly with larger sample sizes and higher-quality data. Meanwhile, the provided detailed host star abundances can be employed to study the composition of the planets within the current sample, thereby contributing to a better understanding of the aforementioned discrepancy.}
  


\titlerunning{TTV -- RV density discrepancy}

\maketitle

\section{Introduction}                                  \label{sec:intro}

More than 5500 exoplanets have been discovered so far\footnote{https://exoplanetarchive.ipac.caltech.edu/}. This has allowed us to perform statistical studies of their properties that provided valuable insights into their formation and evolution \citep[e.g.][]{Adibekyan-19, Zhu-21, Luque-22}. In this respect, a precise, and if possible, accurate determination of the main physical properties of these planets, such as their masses and radii, is critical.

Exoplanetary masses are primarily determined using the radial velocity (RV) method. In some instances, planet-planet interactions can result in transit-timing variations (TTVs), and if these are measured, they can be used to determine the planetary masses \citep[][]{Agol-05}. Interestingly, there is a significant discrepancy between the masses obtained from these two methods for planets with masses up to about that of Neptune \citep[e.g.][]{Mills-17}. Specifically, RV-detected planets appear to be denser than TTV-detected planets of the same radius \citep{Leleu-23}. Although the origins of this discrepancy are not yet fully understood, it is likely influenced by a combination of astrophysical and observational factors \citep{Leleu-23}.

One potential approach for investigating this issue is to consider it from the perspective of the host stars. The composition and density of low-mass planets are connected to the chemical composition of their host stars \citep{Adibekyan-21}. Therefore, a straightforward approach is to compare the abundances of planet-forming elements in the host stars. 

We present the results of our spectroscopic analysis of FGK dwarf stars hosting low-mass TTV- and RV-exoplanets. The paper is structured as follows: In Sect.\ref{sec:sample}, we introduce the sample, and in Sect.\ref{sec:stars}, we detail the composition of the stars. Our findings are discussed in Sect.\ref{sec:results}, and we provide a summary of the work in Sect.\ref{sec:summary}.

\section{The sample}                        \label{sec:sample}

We started our sample selection from the NASA Exoplanet Archive (NEA)\footnote{https://exoplanetarchive.ipac.caltech.edu/} by selecting planets with masses below 10 $M_{\mathrm{\oplus}}$ orbiting FGK-type stars with effective temperatures ($T_{\mathrm{eff}}$) between 5000 and 6500K.  This temperature range allows for the precise determination of stellar atmospheric abundances of relevant elements. We excluded two TTV planets (Kepler-51,c and Kepler-51,d) that have unusually large radii of about 9 $R_{\mathrm{\oplus}}$ \citep{Masuda-14} and were found to be not reliable \citep{Hadden-17}. We selected 62 planets (orbiting 48 stars) with a precision in both mass and radius better than 40\%. For 46 of the host stars (hosting 59 planets), we collected high-resolution optical spectra with a signal-to-noise  ratio (S/N) of at least 40.

\begin{figure}
\begin{center}
\includegraphics[width=0.8\linewidth]{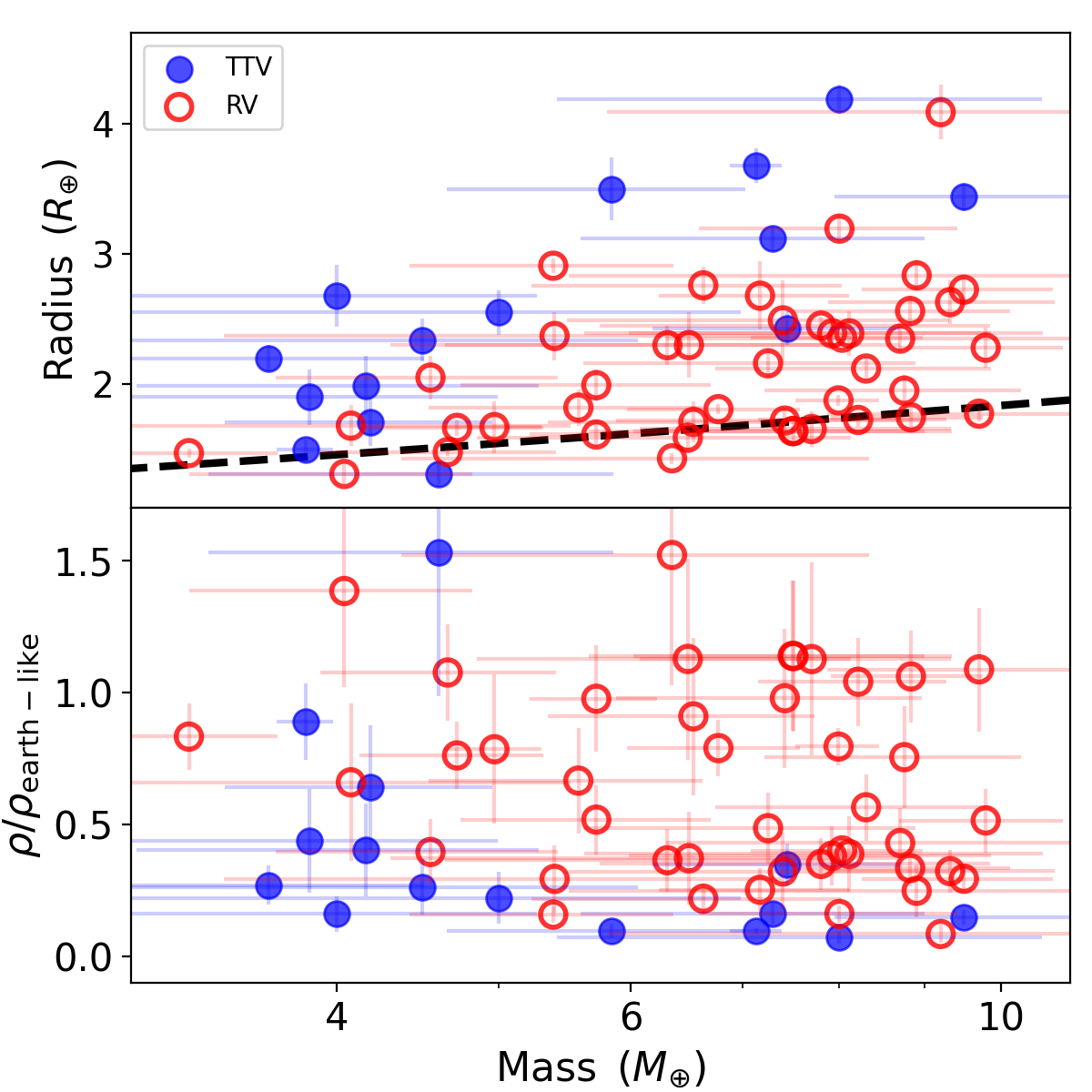}
\end{center}
\vspace{-0.4cm}
\caption{Mass--radius (top) and scaled density--radius(bottom) diagrams for the planets in our sample.  The blue and red symbols represent planets with TTV and RV mass determinations, respectively. The dashed black curve displays the expected mass-radius relation for an Earth-like composition \citep[32\% Fe + 68\% MgSiO3,][]{Dorn-17}.}
\label{fig:mr_full}
\end{figure}

The distribution of 59 planets on the mass-radius diagram is depicted in the top panel of Figure~\ref{fig:mr_full}. The masses of 44 of these planets were determined using the RV method, and the masses of the remaining 15 planets were determined through TTV. We note that planetary masses and radii were determined by different authors, which might introduce a factor contributing to uncertainty. The figure confirms previous findings that TTV planets tend to have larger radii for a given mass when compared to planets whose masses are determined via the RV method.

In the bottom panel of Figure~\ref{fig:mr_full}, we show the density of the planets ($\rho$), normalized to that expected for an Earth-like composition ($\rho_{\mathrm{Earth-like}}$) \citep{Dorn-17}. This normalization takes into account that planets with different masses may have different densities, even when the composition is the same, due to gravitational compression. The figure suggests systematically lower densities for TTV-detected planets when compared with their RV-detected counterparts.

\section{Composition of stars and planet-building blocks}                        \label{sec:stars}

In this section, we outline the method for determining the abundance of stars and the composition of the building blocks for planets.

\subsection{Stellar parameters and abundances}

As mentioned earlier, we used high-resolution spectra of the 46 planet-hosting stars to homogeneously determine their stellar parameters and eventually, their atmospheric compositions. 

We determined the stellar atmospheric parameters ($T_{\mathrm{eff}}$, surface gravity $\log g$, microturbulent velocity $V_{\mathrm{tur}}$, and metallicity [Fe/H]) of the sample stars following our previous works \citep[e.g.][]{Santos-13,Sousa-14}. In our local thermodynamic equilibrium (LTE) analysis, we measured the equivalent widths (EW) of iron lines using the code ARES v2 \footnote{The last version of the ARES code (ARES v2) can be downloaded at http://www.astro.up.pt/$\sim$sousasag/ares} \citep{Sousa-15} and  imposed excitation and ionization equilibrium. We adopted a grid of Kurucz model atmospheres \citep{Kurucz-93} and the 2014 version of the radiative transfer code MOOG \citep{Sneden-73}. The stars of our sample cover the following range of parameters: 5081 $<$ $T_{\mathrm{eff}}$ $<$ 6330 K, -0.47 $<$ [Fe/H] $<$ 0.4 dex, and 4.0 $<$ $\log g$ $<$ 4.5 dex (they are all dwarfs).

The stellar abundances of the elements were also derived using the same tools and models as for the stellar parameter determination, and we used the classical curve-of-growth analysis method assuming LTE. Although the EWs of the spectral lines were automatically measured with ARES, for the elements for which only two or three lines were available, we visually inspected the EWs measurements. To derive the chemical abundances of Mg, Si, Ti, and Ni, we closely followed the methods described in our previous works \citep{Adibekyan-12, Adibekyan-15}. The abundances of the volatile elements O and C were derived following our previous works \citep{Delgado-10, Delgado-21, Bertrandelis-15}. Following the discussion in \citet{Delgado-21}, we opted to use the 6158\AA{} line to determine the abundance of O. The EWs of this spectral line were manually measured using the \texttt{splot} task in IRAF.

It is particularly challenging to determine the C and O abundances for stars cooler than about 5200 K, especially when the S/N falls below approximately 200 \citep{Delgado-10, Bertrandelis-15}. We were unable to determine the C abundance for 17 of the stars and the O abundance for 21 stars. For these stars, we predicted their C and O abundances using a machine-learning (ML) model, as described in Section~\ref{sec:ML}.

\begin{figure}
\begin{center}
\includegraphics[width=1\linewidth]{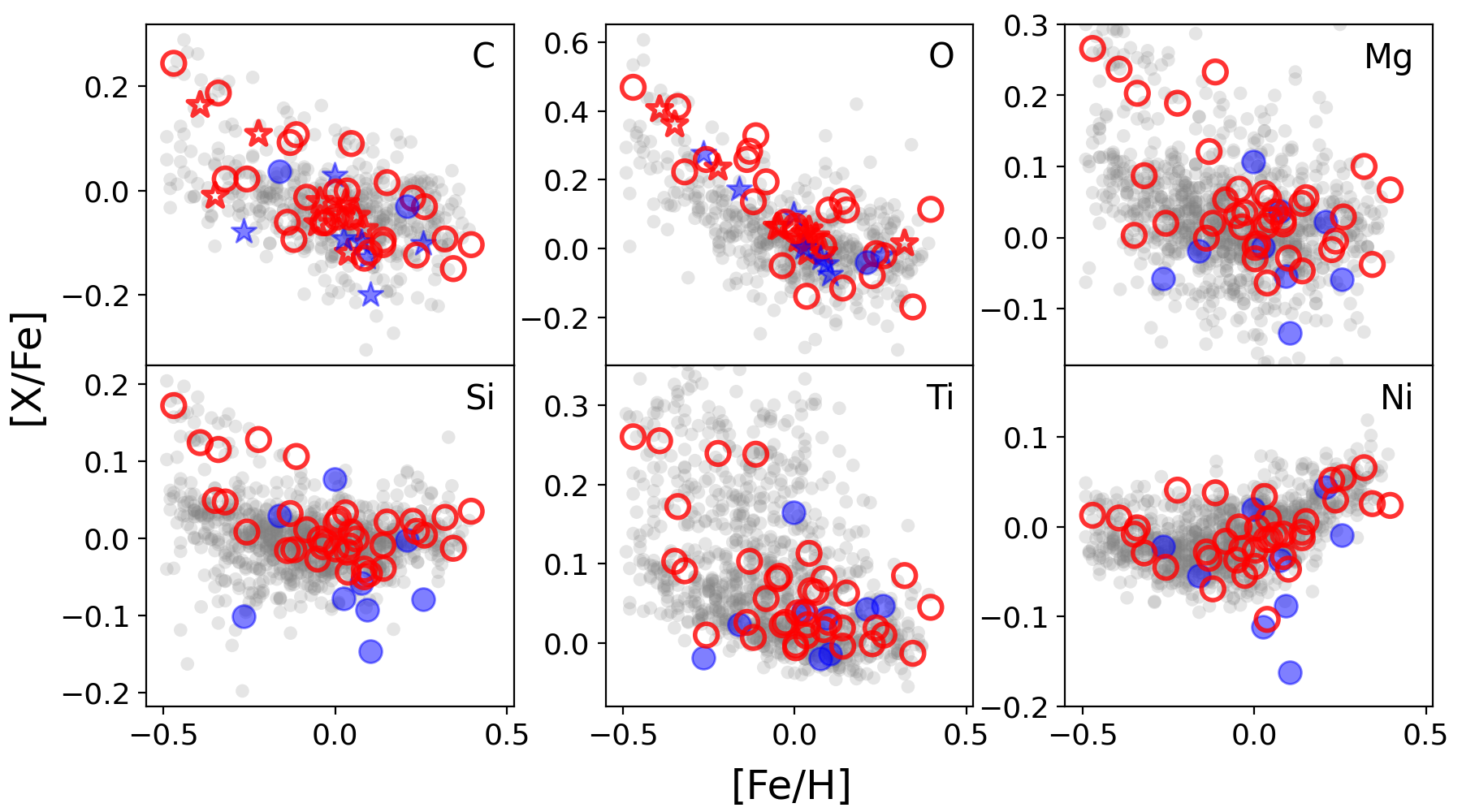}
\end{center}
\vspace{-0.3cm}
\caption{Abundance ratios ([X/Fe]) as a function of metallicity for the sample stars. The blue and red symbols represent planets with TTV and RV mass determinations, respectively. The star symbols indicate the abundances predicted using the ML method, and the circles represent the spectroscopic determinations. The gray dots represent stars without planets from the HARPS GTO sample \citep{Adibekyan-12}.}
\label{fig:el_fe}
\end{figure}

In Fig.~\ref{fig:el_fe}, we display the [X/Fe] abundance ratios as a function of metallicity, where X represents C, O, Mg, Si, Ti, and Ni. The stars hosting TTV-mass and RV-mass planets are represented in blue and red, respectively. The two populations visually appear to exhibit a similar distribution of these elements. However, the Anderson-Darling (AD) and Kolmogorov-Smirnov (KS) two-sample tests suggest that the distributions of the [Mg/Fe] and [Si/Fe] abundance ratios do not originate from the same parent distribution (p-values $<$ 0.05). These differences may play a role in the interior structure and composition of low-mass TTV and RV planets, as we discuss in the following sections.

\subsection{Composition of the planet-building blocks}

Taking the abundances of C, O, Mg, Si, and Fe into account, we estimated the iron-to-silicate mass fraction ($f_{\mathrm{iron}}$) and the water-mass fraction ($w_{f}$) of the protoplanetary disks using the simple stoichiometric model presented in \citet{Santos-15, Santos-17}. While the determination of $f_{\mathrm{iron}}$ depends solely on the abundances of Mg, Si, and Fe, the determination of $w_{f}$ relies on the abundances of all five elements.

\begin{figure}
\begin{center}
\includegraphics[width=1\linewidth]{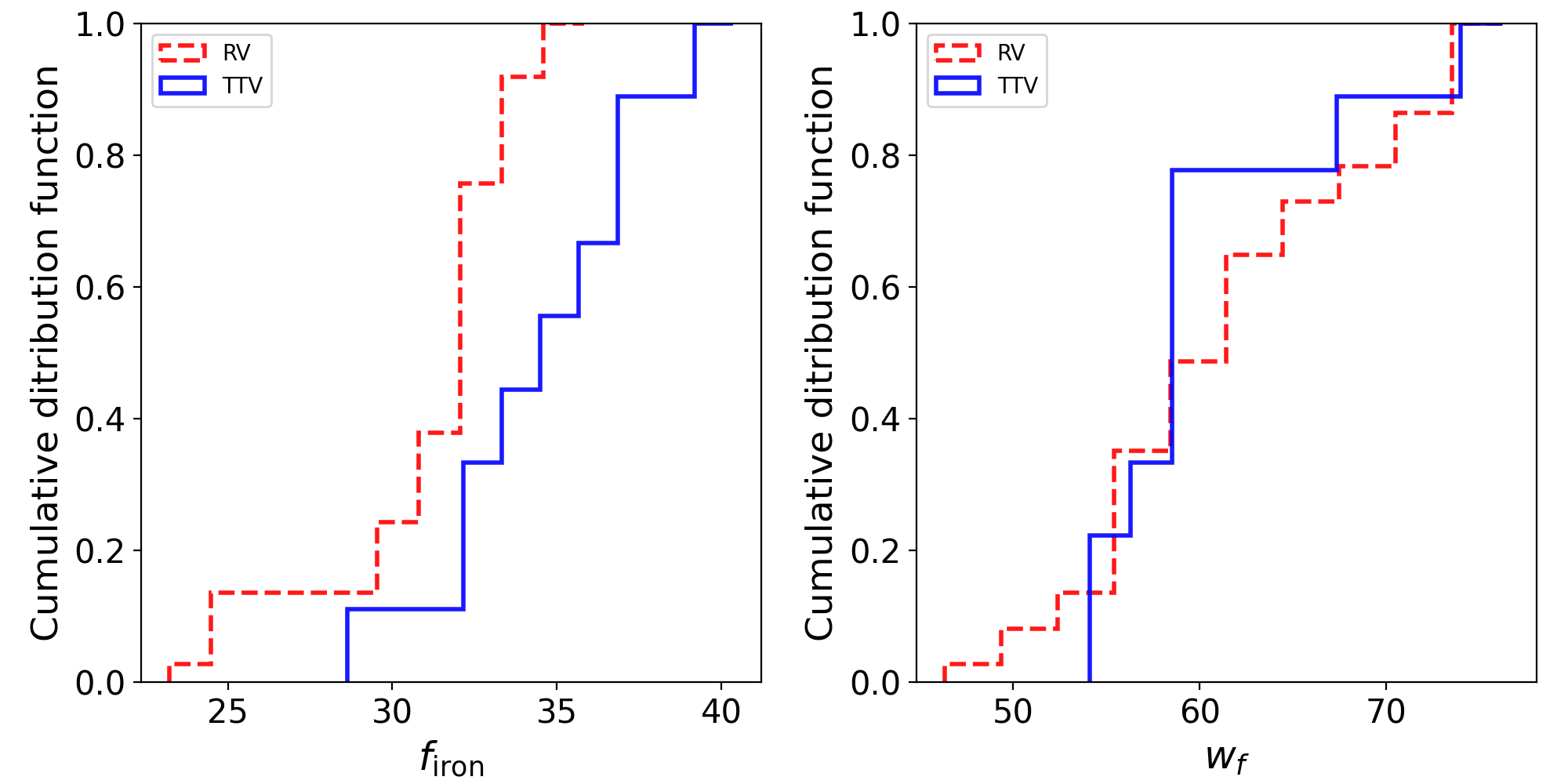}
\end{center}
\vspace{-0.4cm}
\caption{Cumulative distributions of $f_{\mathrm{iron}}$ and $w_{f}$ for the stars hosting planets with TTV and RV mass determinations.}
\label{fig:f_iron}
\end{figure}

In Fig.~\ref{fig:f_iron}, we show the cumulative distributions of $f_{\mathrm{iron}}$ and $w_{f}$ for stars hosting TTV-mass and RV-mass planets. Both the two-sample AD  and KS tests yield a p-value below 0.02, suggesting that the two samples have distinct distributions of $f_{\mathrm{iron}}$. The p-values of the AD and KS tests for  $w_{f}$ indicate that the distributions might come from the same parent distribution.

\section{Results}                        \label{sec:results}

\begin{figure}
\begin{center}
\includegraphics[width=1\linewidth]{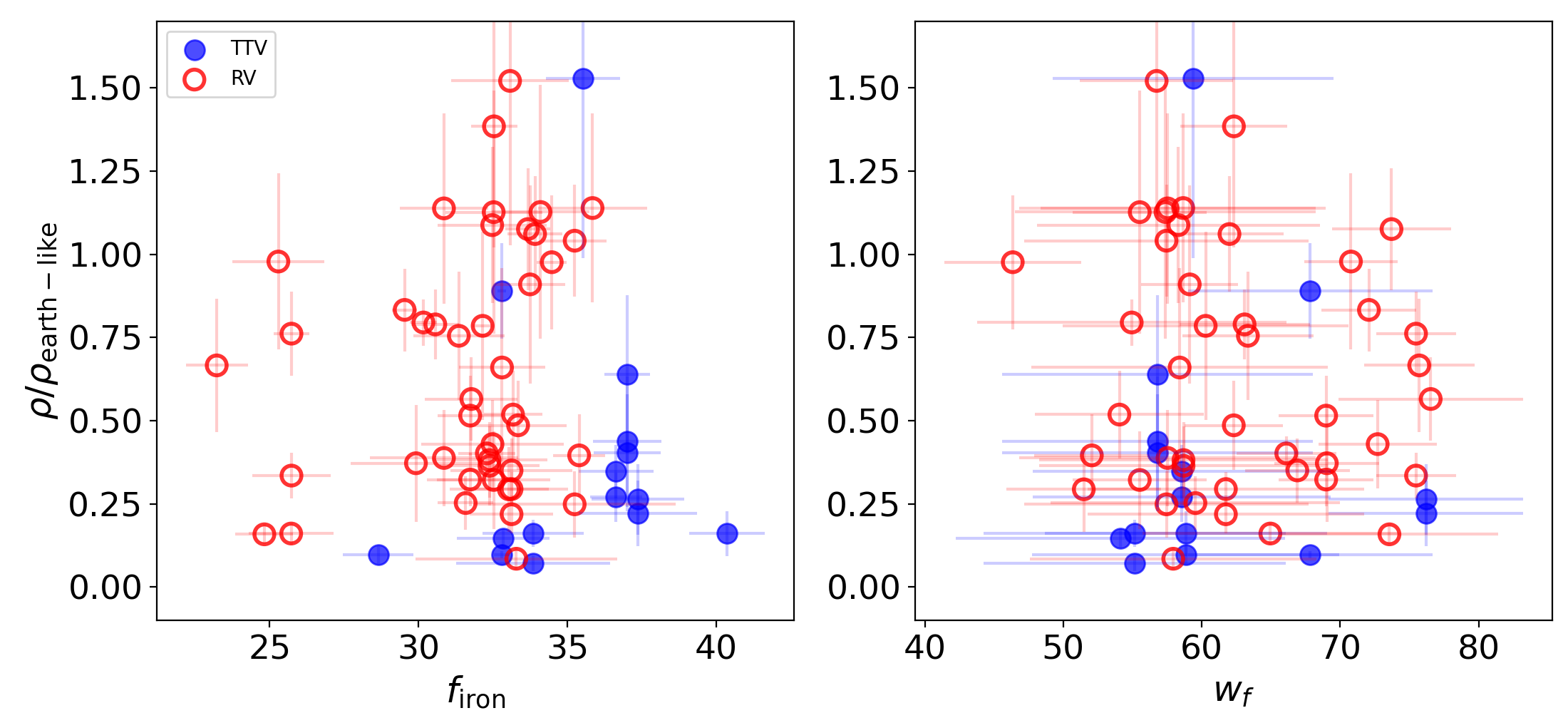}
\end{center}
\vspace{-0.4cm}
\caption{Scaled density of planets as a function of $f_{\mathrm{iron}}$ and $w_{f}$.}
\label{fig:density_fw_fraction}
\end{figure}

Under the assumption that low-mass ($<$ 10 $M_{\mathrm{\oplus}}$) and small ($<$ 2 $R_{\mathrm{\oplus}}$) planets are composed of metallic cores and mantles, \citet{Adibekyan-21} observed a correlation between the scaled density of these planets and their $f_{\mathrm{iron}}$ content. In Fig.~\ref{fig:density_fw_fraction}, we show the dependence of the scaled density of planets on their $f_{\mathrm{iron}}$ and $w_{f}$. The figure does not reveal a clear correlation between these parameters. This lack of a correlation arises because many of the planets in the current sample are larger than 2 $R_{\mathrm{\oplus}}$, with non-negligible atmospheres\footnote{According to \citet{Zeng-19}, in an extreme case, if a planet with a mass of 5 $M_{\mathrm{\oplus}}$ is composed entirely of water, its radius would be approximately 2 $R_{\mathrm{\oplus}}$, which is smaller than the radii of many planets in the current sample.}. In this scenario, the size of the atmospheres, rather than the interior composition, influences the radius, and consequently, the scaled density of the planets. 

\begin{figure}
\begin{center}
\includegraphics[width=0.8\linewidth]{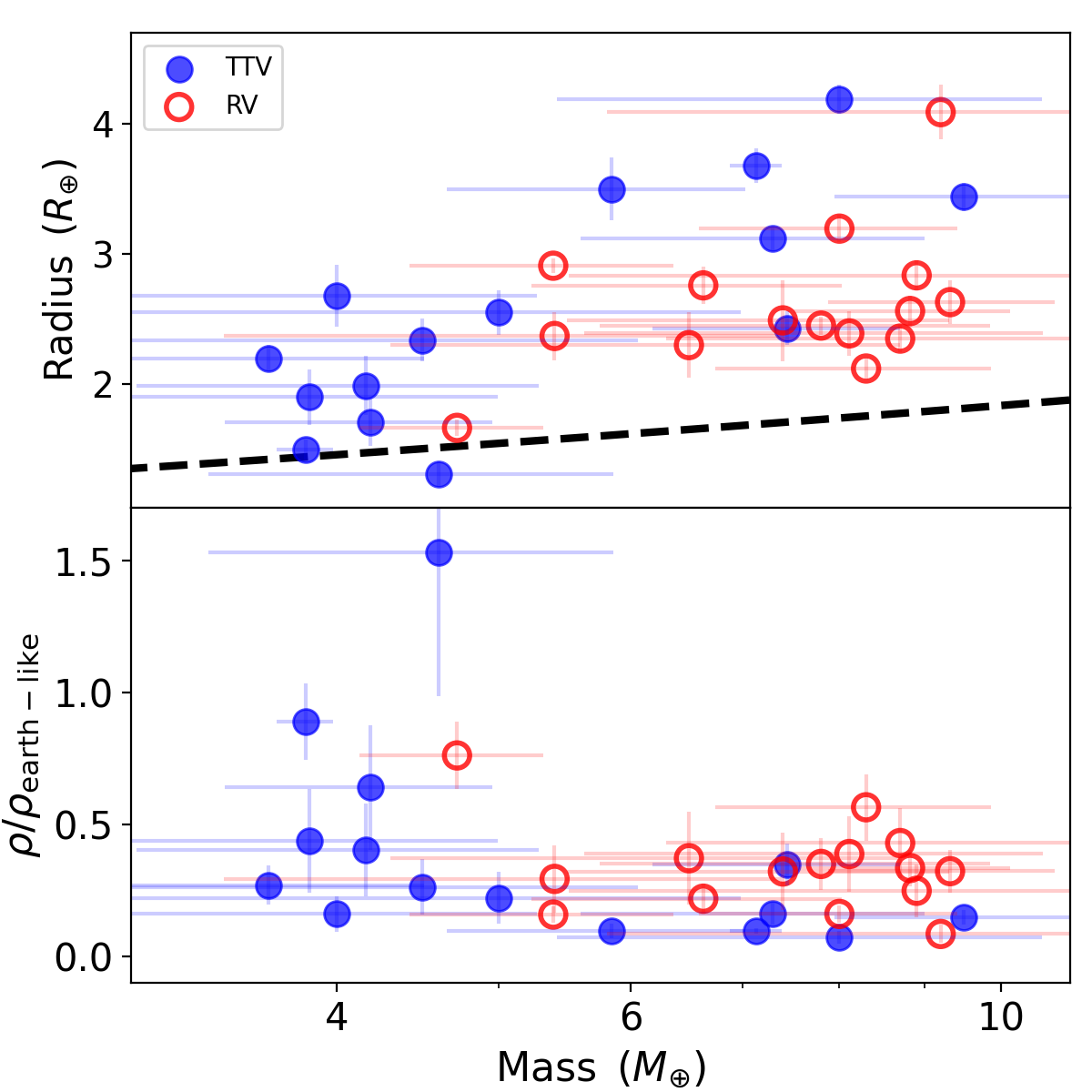}
\end{center}
\vspace{-0.5cm}
\caption{Same as Fig.~\ref{fig:mr_full}, but for planets with orbital periods longer than 7 days.}
\label{fig:mr_temperate}
\end{figure}

\begin{figure}
\begin{center}
\includegraphics[width=1\linewidth]{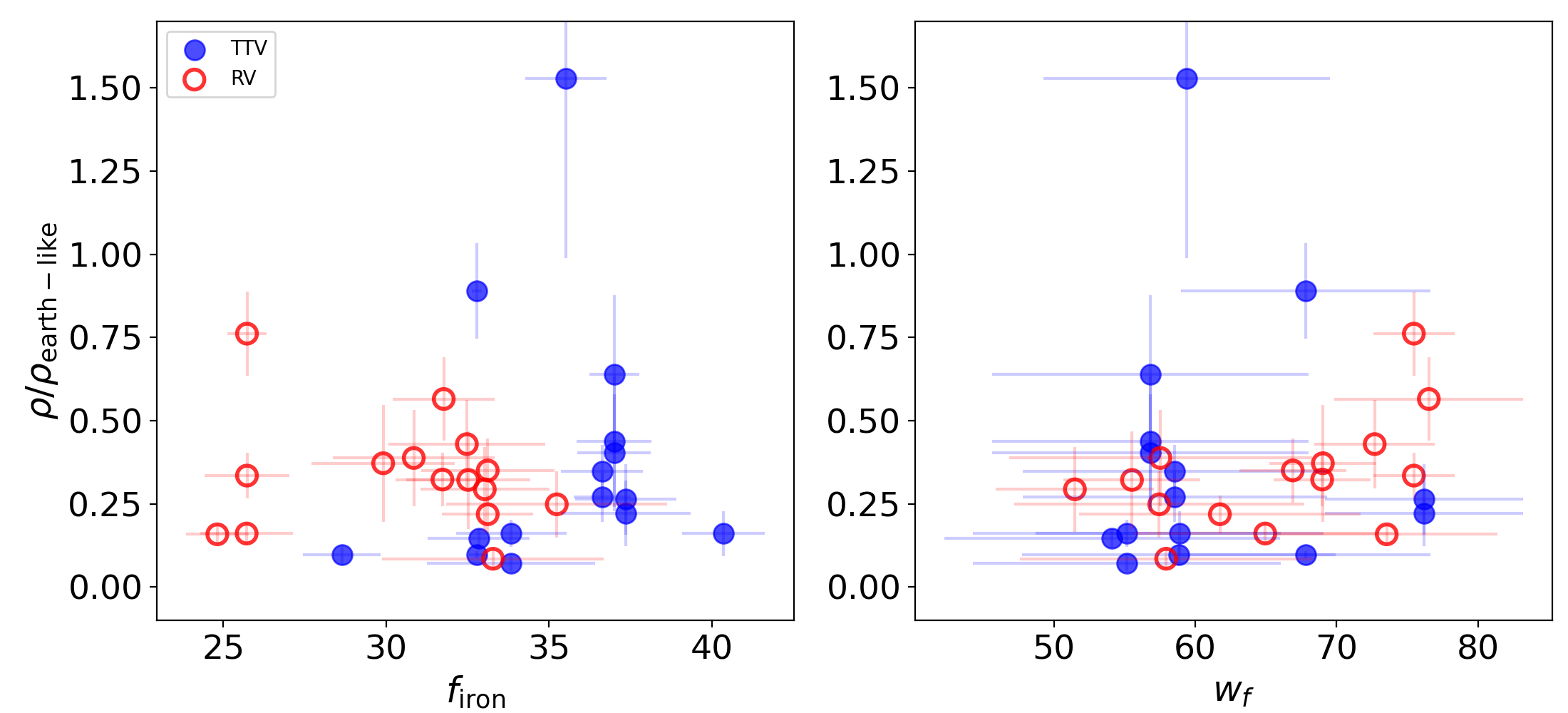}
\end{center}
\vspace{-0.5cm}
\caption{Same as Fig.~\ref{fig:density_fw_fraction}, but for planets with orbital periods longer than 7 days.}
\label{fig:density_fw_fraction_temperate}
\end{figure}

The size of the atmospheres, among other factors, depends on the proximity of the planets to their stars. In the current sample, all TTV planets have orbital periods longer than 7 days, while 29 out of 44 RV planets have shorter periods. When considering only the long-period planets (see Figs.~\ref{fig:mr_temperate} and ~\ref{fig:density_fw_fraction_temperate}), we find that primarily the planets with low scaled densities remain. Of these, the TTV planet hosts generally exhibit higher values of $f_{\mathrm{iron}}$ than the RV planet hosts: Their mean value is 35.3$\pm$2.8($\pm$0.7\footnote{The values in parentheses are standard errors of the mean, calculated as the standard deviation divided by the square root of the sample size.}), and 30.6$\pm$3.4($\pm$0.9), respectively. The p-values of the AD and KS tests return values lower than 0.001. The TTV planet hosts also seem to have slightly lower values of $w_{f}$ (61.1$\pm$7.3($\pm$1.9) and 65.7$\pm$8.3($\pm$2.1)), although the AD and KS tests indicate that the difference is not statistically significant.

\begin{figure}
\begin{center}
\includegraphics[width=1\linewidth]{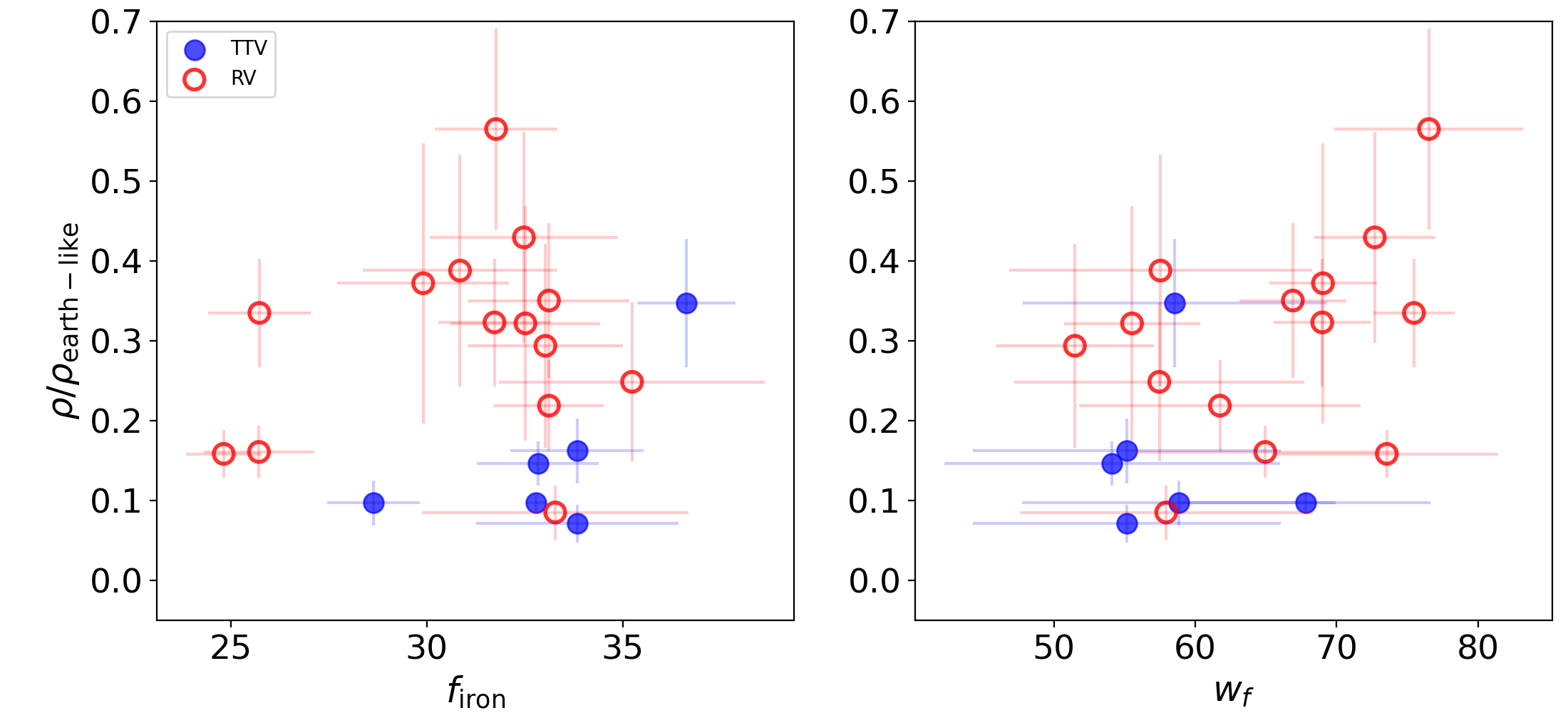}
\end{center}
\vspace{-0.5cm}
\caption{Same as Fig.~\ref{fig:density_fw_fraction}, but for planets with orbital periods longer than 7 days and masses higher than 5 $M_{\mathrm{\oplus}}$.}
\label{fig:density_fw_fraction_temperate_massive}
\end{figure}

The restriction on orbital period eliminates most RV planets with masses below 5 $M_{\mathrm{\oplus}}$ because at relatively wide orbits, only high-mass planets can be detected using the RV method. When applying a lower mass limit cut at 5 $M_{\mathrm{\oplus}}$, the two samples display a relatively similar mass distributions. However, the sample sizes decrease to only 6 TTV planets (orbiting 5 stars) and 14 RV planets (orbiting 14 stars). Fig.~\ref{fig:density_fw_fraction_temperate_massive} indicates at higher $f_{\mathrm{iron}}$ (33.1$\pm$2.6($\pm$1.1) and 30.9$\pm$3.2($\pm$0.9)) and lower $w_{f}$ (58.3$\pm$5.0($\pm$2.1) and 64.9$\pm$8.1($\pm$2.1)) for the TTV planets than for their RV counterparts. Nonetheless, the small sample sizes and large uncertainties make a firm conclusion challenging.

\begin{figure}
\begin{center}
\includegraphics[width=1\linewidth]{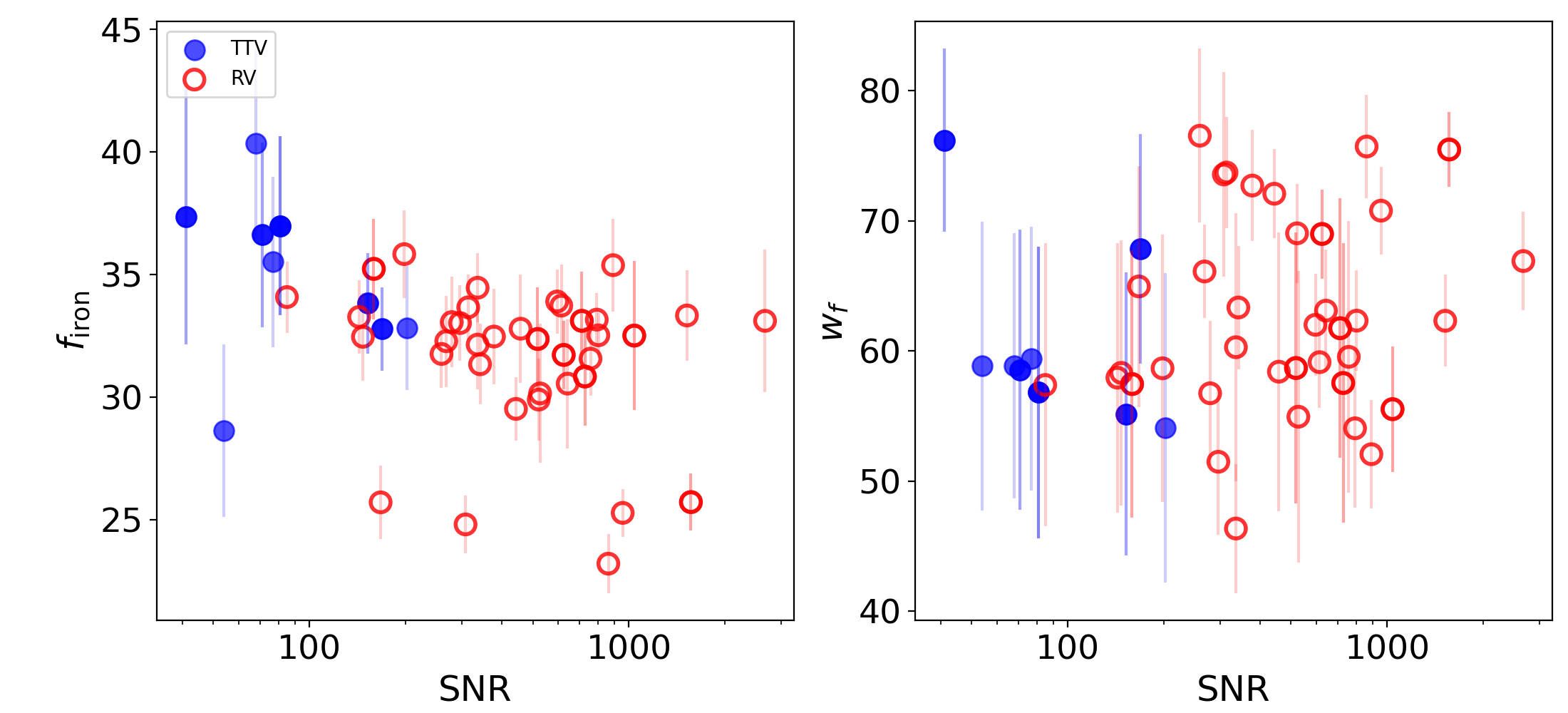}
\end{center}
\vspace{-0.5cm}
\caption{Dependence of $f_{\mathrm{iron}}$ and $w_{f}$ on SNR.}
\label{fig:fw_fraction_snr}
\end{figure}

\begin{figure}
\begin{center}
\includegraphics[width=1\linewidth]{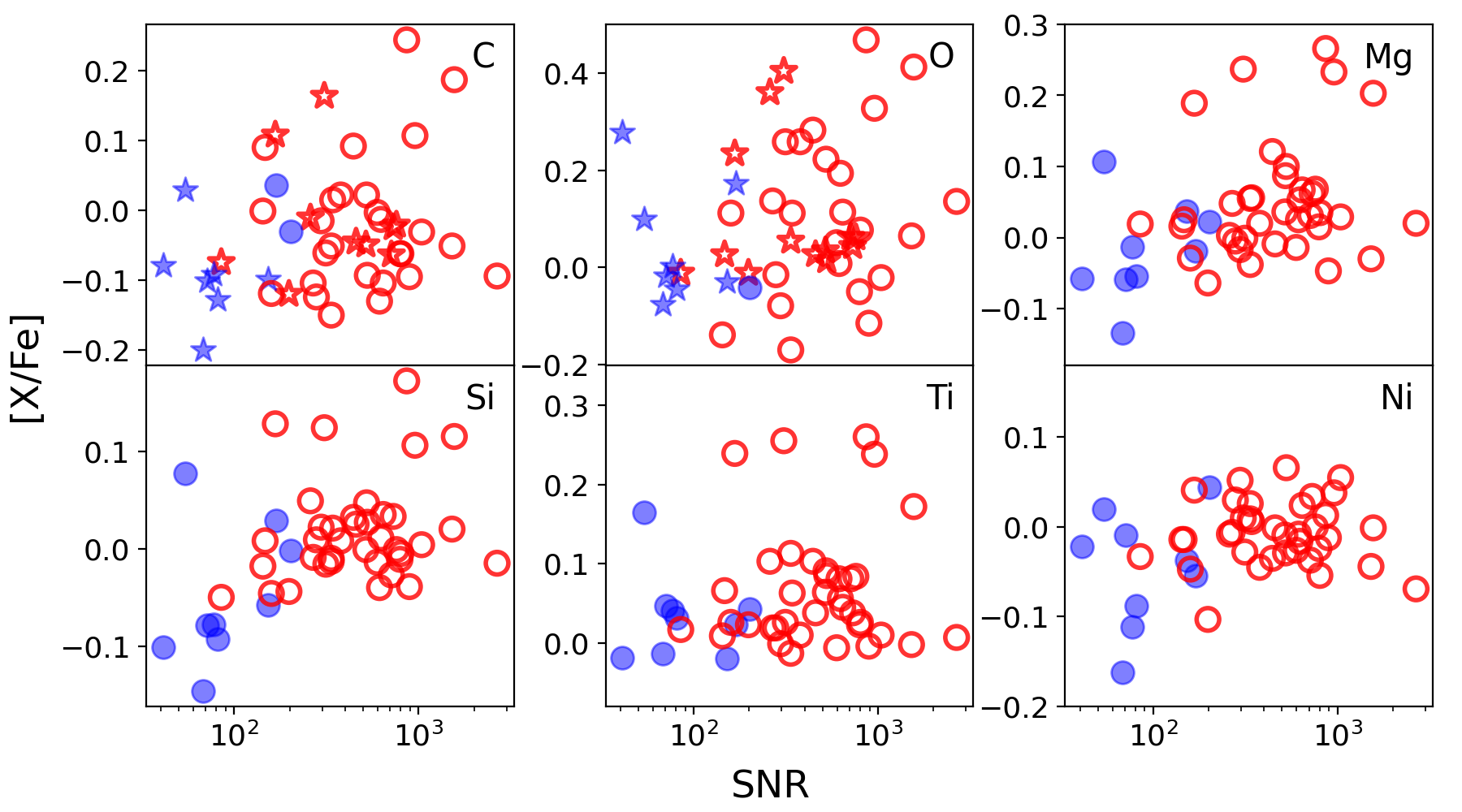}
\end{center}
\vspace{-0.5cm}
\caption{Dependence of the abundance ratios on SNR.}
\label{fig:abund_snr}
\end{figure}

A word of caution should be added here. Fig.\ref{fig:fw_fraction_snr} appears to suggest a relation between $f_{\mathrm{iron}}$ and SNR, at least for values below about 100. Unfortunately, the spectra of most TTV planet-hosting stars are affected by this\footnote{To determine masses with the RV method, a relatively large number of spectra is required, which, when combined, result in a high SNR.}, making it difficult to determine whether the lowest values of $f_{\mathrm{iron}}$ arise from a low SNR or are caused by the chemically distinct stars that these are TTV planets orbit. In Fig.\ref{fig:abund_snr}, we show the dependence of the abundance ratios on SNR. The figure shows a slight dependence of [Si/Fe] on SNR. Stars with low SNR spectra also exhibit low $f_{\mathrm{iron}}$. To quantify the relation between the abundance ratios and SNR, we computed the Pearson correlation coefficients and their corresponding p-values. The highest correlation coefficient, 0.27, is observed for [Si/Fe], with a p-value of 0.07, indicating that it is not statistically significant. The Si abundances were derived using 14 spectral lines that cover a wide range of wavelengths, making it hard to conclude that a low SNR has a high systematic impact on the abundance determination of this element. In addition, our visiual inspection of the Gaussian fits of the spectral lines of Si did not reveal any indication of incorrect or suspicious measurement. 

We also tested whether the stars hosting TTV and RV planets have different kinematic properties or belong to different galactic populations (see Sect.\ref{sec:kinematics}). In Fig.\ref{fig:kinematics}, we display the distributions of the planet hosts in velocity space. The figure, as well as our tests, revealed no differences in their kinematic properties. For both samples, approximately one out of ten stars belongs to the Galactic thick-disk population, and the rest belong to the thin disk, which is typical for field stars in the solar neighborhood \citep[e.g.,][]{Adibekyan-13}.

\section{Summary}                        \label{sec:summary}

We employed the largest dataset of FGK-type stars hosting planets with massed determined via TTV and RV methods. We determined the host star abundances, and, employing a simple stoichiometric model \citep{Santos-15, Santos-17}, we estimated the iron-to-silicate mass fraction ($f_{\mathrm{iron}}$) and the water-mass fraction ($w_{f}$) of the protoplanetary disks. We then explored the possibility that the discrepant mass/densities in these two populations \citep[e.g.][]{Leleu-23} might be linked to the composition of the protoplanetary disks in which they formed.

Our tests show indications that TTV planets tend to have higher $f_{\mathrm{iron}}$ and lower $w_{f}$ than RV planets. However, these results should be interpreted with caution due to differences in planetary properties (e.g., orbital periods for a given mass) between the two samples and the relatively small sample sizes. Furthermore, the spectra of stars hosting TTV planets exhibit a significantly lower SNR than their RV counterparts, which could influence the results.

Even when these limitations of the analysis are ignored, it is challenging to interpret the results. In particular, the higher $f_{\mathrm{iron}}$ and lower $w_{f}$ observed in TTV planets would suggest that these planets (without considering the atmospheres) are denser than their RV counterparts on average. This would imply that larger atmospheres are required on TTV planets to account for their lower overall densities.  To the best of our knowledge, no previous studies have suggested such a correlation. 

It is also possible that the specific composition (e.g., high $f_{\mathrm{iron}}$)  of the disk might facilitate the formation of compact multiplanet systems (e.g., TTV planets), suggesting that the density could be related to the formation and evolution history of these systems \citep[e.g.][]{Lee-16, Izidoro-17}.

In conclusion, investigating the TTV-RV mass discrepancy through a host star chemical analysis clearly holds promise for future research. However, to make more conclusive determinations, larger sample sizes and higher-quality data are required. In the meantime, we hope that the provided host star abundances can be used by modelers of the exoplanet interior to study the composition of both TTV and RV planets, which may help to shed light on this intriguing problem.

\begin{acknowledgements}
This work was supported by Funda\c{c}\~ao para a Ci\^encia e Tecnologia through national funds and by FEDER through COMPETE2020 - Programa Operacional Competitividade e Internacionalização by these grants: UIDB/04434/2020; UIDP/04434/2020; 2022.06962.PTDC. Funded/Co-funded by the European Union (ERC, FIERCE, 101052347). Views and opinions expressed are however those of the author(s) only and do not necessarily reflect those of the European Union or the European Research Council. Neither the European Union nor the granting authority can be held responsible for them.

This research has made use of the NASA Exoplanet Archive operated by the California Institute of Technology under contract with the National Aeronautics and Space Administration under the Exoplanet Exploration Program.

In this work we used the Python language and several scientific packages: Numpy \citep{van_der_Walt-11}, Scipy \citep{Virtanen-20}, Pandas \citep{mckinney-proc-scipy-2010}, and Matplotlib \citep{Hunter-07}.
\end{acknowledgements}

\bibliographystyle{aa}
\bibliography{references}


\begin{appendix}

\section{Prediction of C and O abundances with XGBoost}        \label{sec:ML}

We were unable to determine the spectroscopic abundances of carbon (C) and oxygen (O) for 17 and 21 stars, respectively. For these stars, we empirically estimated the C and O abundances using the machine-learning algorithm called 'XGBRegressor'  \citep{Chen-16}. The estimation of C and O (the target variables) was based on the abundance of iron and the mean abundance of Mg, Si, and Ti (the input feature variables). While using any of these abundances individually or all of them simultaneously provides similar results, taking the mean of the abundances slightly improves the performance of the model predictions.

Our initial sample was based on the HARPS sample \cite{Adibekyan-12}. C abundance of 758 stars and O abundance of 610 stars were taken from \citet{Delgado-21}, and the iron abundance was taken from \citet{Delgado-17}. These samples were divided into training (70\%), validation (20\%), and testing (10\%) datasets. We used the Optuna package for an automatic hyperparameter optimization \citep{Akiba-19}. The mean absolute error for the estimated C and O abundances on the test dataset is 0.06 and 0.07 dex, respectively. We note that this is consistent with the mean uncertainties in the abundance determinations of these elements (0.05 dex for C and 0.07 dex for O).

\begin{figure}
\begin{center}
\includegraphics[width=1\linewidth]{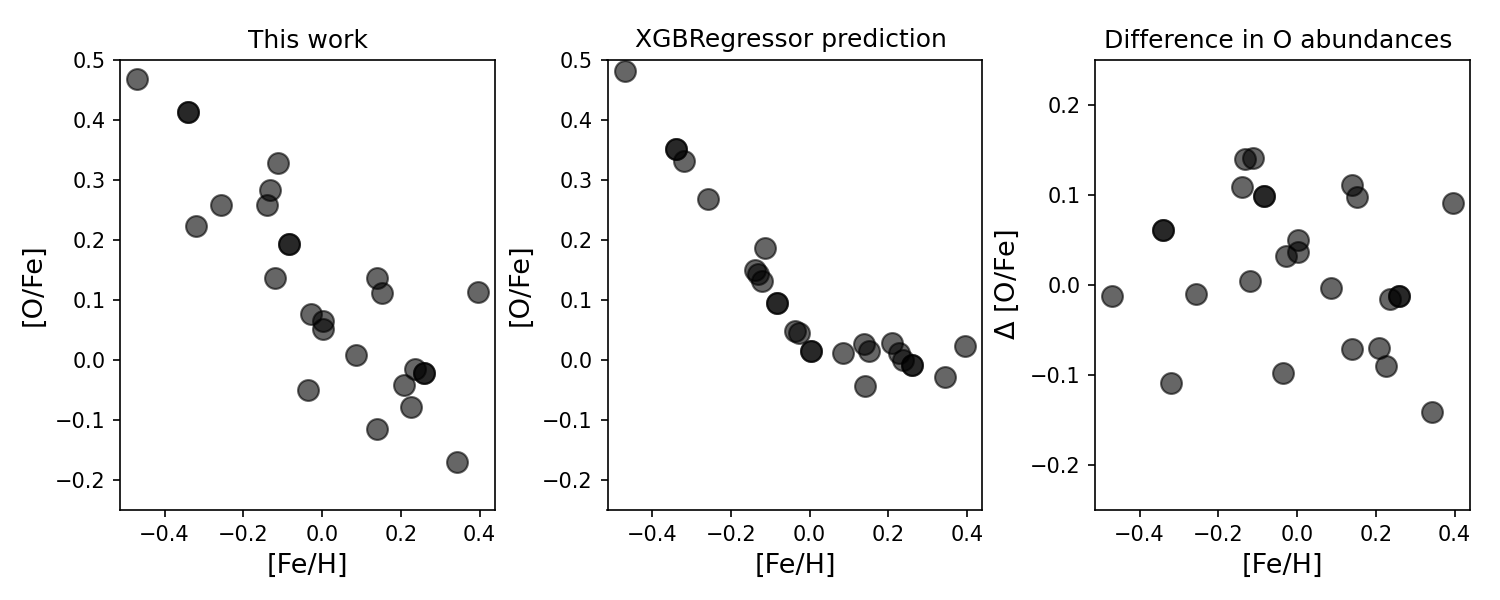}

\includegraphics[width=1\linewidth]{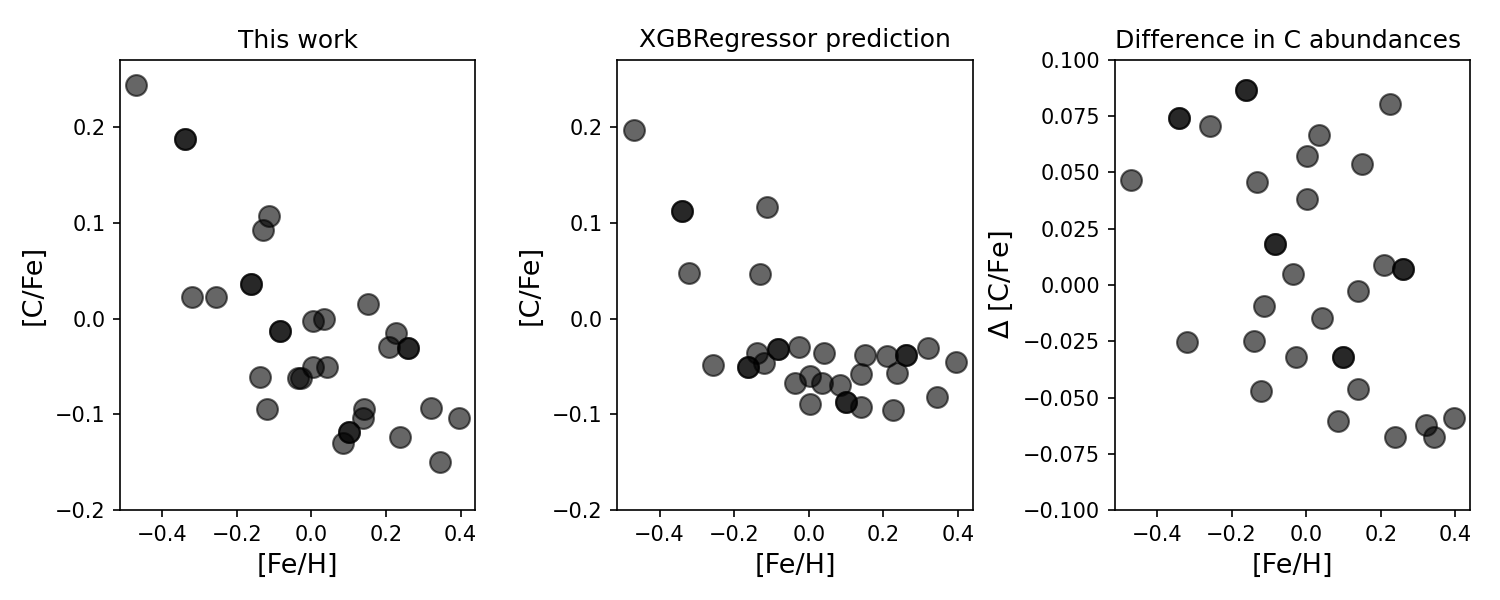}
\end{center}
\caption{C and O abundances as a function of metallicity of the sample stars that we spectroscopically determined (leftmost panels), predicted using the model (middle panels), and the difference in the two estimates (rightmost panels).}
\label{fig:c_o_predict}
\end{figure}

We also tested the estimated C and O abundances for the current sample of planet-hosting stars, for which we determined the abundances of these elements (see Fig.~\ref{fig:c_o_predict}). The mean difference and standard deviation for C and O are 0.01$\pm$0.05 and 0.01$\pm$0.08 dex, respectively.

\newpage
\section{Kinematics of the host stars}        \label{sec:kinematics}

We computed the Galactic space velocity ($U,V,W$) of the planet-hosting stars based on their coordinates, proper motion, parallax, and systemic RV extracted from Gaia DR3, following the formulation developed in \citet{Johnson-87}. We considered the solar peculiar motion from \citet{Schonrich-10} to calculate the velocities with respect to the local standard of rest. These velocities were used to estimate the probability that the stars belong to the thin disk, the thick disk, or the halo, based on the kinematic approach presented in \citet{Adibekyan-12}\footnote{We used the \texttt{GalVel\_Pop} package available at \url{https://github.com/vadibekyan/GalVel_Pop}}.

\begin{figure}
\begin{center}
\includegraphics[width=1\linewidth]{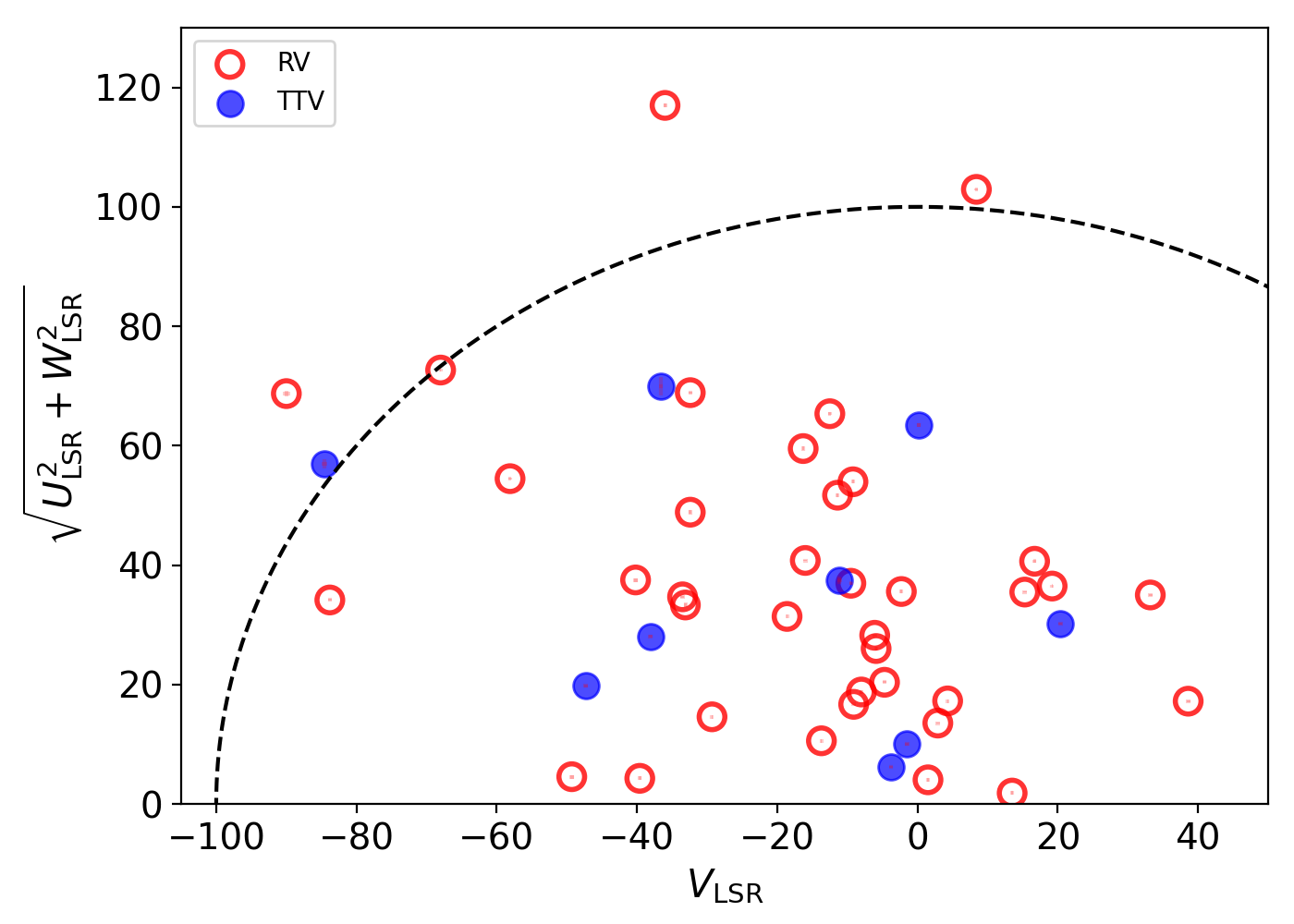}
\end{center}
\vspace{-0.5cm}
\caption{Toomre diagram for the full sample. The typical uncertainties are below 1 km/s and are hardly visible in the plot. The dotted line shows constant values of the total Galactic velocity $V_{\mathrm{tot}}$ = 100 km/s.
}
\label{fig:kinematics}
\end{figure}

\end{appendix}

\end{document}